\documentclass[jkps,twoside,twocolumn,showpacs,showkeys]{revtex4}
\newcommand{\sbar}[1]{#1\!\!\!/}
\usepackage{graphicx}
\usepackage{indentfirst}
\usepackage{amssymb}
\usepackage{amsmath}
\usepackage{simplewick}

\begin{document}
\setcounter{page}{1}
\title[]{Pair Production of Majorana Neutrinos by Annihilation of Charged Particles in  High Energy  Collision
}

\author{Y. M. \surname{Goh}}
\email{youngmoon@hep.hanyang.ac.kr}
\author{H. K. \surname{Lee}}
\email{hyunkyu@hanyang.ac.kr}
\author{W.-G. \surname{Paeng}}
\email{wgpaeng0@hanyang.ac.kr}
\author{Y. \surname{Yoon}}
\email{cem@hanyang.ac.kr}
\affiliation{Department of Physics, Hanyang University, Seoul 133-791}
\received{}

\begin{abstract}
   Assuming that neutrinos have  non-vanishing magnetic moments, we discuss the possibility of pair production through annihilation of charged fermions in high-energy collisions. Adopting the Pauli interaction for photon-neutrino coupling, we calculate the neutrino pair production cross section in the photon channel and compare the result with  the standard model in $ Z^{0} $ channel. we demonstrated that the enhancement of the production rate for  Majorana neutrino pairs  over the standard model rate can be possible at the   center-of-mass energy of $10 - 100$ TeV for the Large Hadron Collider or the ultra-high-energy cosmic Ray when  the transition magnetic moment is not smaller than $10^{-9} - 10^{-10} \mu_B$.
\end{abstract}


\maketitle

\section{INTRODUCTION}

Atmospheric and  solar neutrino observations \cite{ashie,aharmim},  as well
as reactor experiments \cite{Araki}, provide us strong evidence of the oscillation between
different flavors of neutrinos, which can't be possible if neutrinos are  massless.
Neutrinos are   massive, electrically
neutral fermions with spin 1/2.  Although they are neutral fermions, it has been an interesting  question  how  neutrinos can couple to photons.  One of the possibilities is
the non-vanishing magnetic moment of neutrinos, which  can induce a spin-dependent coupling to photons.

Experimental bounds for the neutrino magnetic moments have been obtained from
experiments for the solar neutrinos, accelerator neutrinos and
reactor neutrinos \cite{wong}.  The upper bound of the neutrino magnetic moment is
found to be in the range of $10^{-10} - 10^{-7}$ $\mu_B$.
Here, $\mu_B$ is the Bohr magneton. A model-dependent bound can also be obtained, for example, in big-bang
nucleosynthesis and SN87a, to be $10^{-12} - 10^{-10}$ $\mu_B$ \cite{gg, elmfors, ag}. The
theoretical bounds \cite{bell} were also discussed recently to get
a rather wider range of upper bounds, $10^{-15} - 10^{-7}$ $\mu_B$.
  In the standard model, the neutrino magnetic moment induced by the one-loop effect \cite{fujikawa}
is $
\mu_{\nu} = 3 \times
10^{-19} \left(\frac{m_{\nu}}{eV}\right) \mu_{B}$, which is
much smaller than the above bounds.

In this work, we take the magnetic moment as a parameter, which might probe the physics beyond the standard model.
 It is interesting to  note that  the effect of the magnetic moments of neutrinos on the
vacuum instability  has recently been investigated in the presence of a strong
external magnetic field \cite{yoon} to  find out that with non-vanishing magnetic moment the vacuum instability appears beyond
the critical field strength, $B_c = \frac{m_{\nu}}{\mu_{\nu}}$, against the pair production of
neutrinos.

We consider a process of charged
fermion-antifermion annihilation into neutrino pairs.  The standard
process is a pair production through the $Z^0$ channel. If neutrinos
have non-vanishing  magnetic moments, they can also be produced in
the photon channel through the Pauli interaction \cite{pauli} on top of the standard process.
Previously, the production of a massive neutrino through a magnetic interaction
has been  calculated and discussed for obtaining the experimental bounds \cite{barut, schgal, deshpande, hklyoon}.

In this work, we calculate the differential  and the total cross
sections for pair production of Majorana neutrinos  with transition magnetic moments in the photon channel through the Pauli interaction. The Majorana neutrino is known to have only a transition magnetic moment, which implies that  the lepton flavor number  is not conserved: a pair produced  via the transition magnetic moment consists of two different lepton flavors.  We consider  an  extreme process
with ultra-high energy, $E_{CM} > 10$ TeV , which is possible for the
hadronic collision in the Large Hadron Collider (LHC) and $E^{GZK}_{CM} \sim 100$ TeV \cite{note}  in the ultra-high-energy cosmic ray (UHECR).
Because of the momentum-dependent coupling in the Pauli interaction, the total cross section
in the photon channel becomes constant in the high-energy region,
$E \gg m_i $, while the cross section in the $Z^0$ channel decreases
as the energy scale increases. Hence, there is a critical energy
beyond which the Pauli interaction dominates the SM process.  We demonstrate that the critical energy can be in the energy region of the LHC or the UHECR provided the transition magnetic moment is not much smaller than
$10^{-9} - 10^{-10}$ $\mu_B$.
It
is also found that the angular distribution of the differential
cross section in the Pauli interaction  has a maximum at $\theta =
\pi/2$ in the center-of-mass system.

In Section II,  The basic feature of the Pauli interaction and the cross section in the center-of-mass system are discussed. In section III, the cross section for Majorana neutrino pairs is discussed in detail, and discussions are given in Section IV.

\section{Pauli interaction}
In relativistic quantum theory, the standard picture is that the motion of a charged fermion is governed by the Dirac equation with an interaction with the external electromagnetic field, $A^{\mu}$, in a
 gauge-covariant way:
\begin{eqnarray}
i\hbar \frac{\partial \psi}{\partial t}=\left[c \boldsymbol{ \alpha } \cdot \left(\boldsymbol{p}-\frac{e}{c}\boldsymbol{A}\right)
+ \beta m c^{2} +e \Phi \right] \psi.
\end{eqnarray}
Pauli  introduced a new  form of the interaction with a magnetic moment \cite{pauli}:
\begin{eqnarray}
\mathcal{L}^{Pauli}= l \frac{1}{2} \psi^{\dagger} \gamma^{\mu}\gamma^{\nu} \psi F_{\mu \nu},
\end{eqnarray}
where the $F_{\mu \nu}$ are the external field strengths in natural units and $l$ has the dimension of
a length.   The corresponding Dirac-Pauli Lagrangian is  given by
\begin{eqnarray}
{\cal L} =
\bar{\psi}(\sbar{p}+\frac{\mu}{2}\sigma^{\mu\nu}F_{\mu\nu}-m)\psi,\label{pauli}
\end{eqnarray}
where $\sigma^{\mu\nu}=\frac{i}{2}[\gamma^{\mu},\gamma^{\nu}]$ and
$g_{\mu\nu}=(+,-,-,-)$.  $\mu$ in the Pauli term measures the
magnitude of the magnetic moment of the neutral fermion.
This Lagrangian describes the interaction of the neutral fermion, but with a non-vanishing magnetic
moment coupled with the external electromagnetic field through the Pauli interaction.
The quantum mechanics with the Pauli interaction has been investigated for various types of electromagnetic fields \cite{lin}.
The vacuum instability against the pair production rate of neutral fermions in linear magnetic fields through the Pauli interaction has been
calculated \cite{yoon}.

Neutrinos do not have electric charges, but they are found to have a non-vanishing mass. Also, it is natural to ask about the possibility of magnetic moments.
However, it is not clear so far  whether they have  non-vanishing magnetic moments through which they can interact with photons directly. We assume, in this work,  the case where the massive neutrinos have non-vanishing magnetic moments or transition magnetic moments.
Then, we can consider  a process in which the neutrino pairs can be produced through photon exchange,
in addition to the weak process by $Z_0$ boson exchange.

In this work, we calculate the cross section of neutrino pair production through the Pauli interaction to investigate the observational effect of the magnetic moment of neutrinos.  We first consider a process in which neutrino pairs are  produced by the annihilation
 of a charged fermion, $q$, with charge $Q e$,
 which has minimal coupling to a photon, lepton, or parton in the hadron collision. In high-energy hadron collisions, the charged fermions can be considered to be  those of partons in hadrons.

The  process we are considering is  the  annihilation of charged
fermions $q$ and $\bar{q}$  into a neutrino through the photon channel with the Pauli interaction.   The interaction Lagrangian for neutrino pair production is
\begin{eqnarray}
\mathcal{L}_{int} =  \frac{\mu}{2}  \bar{\psi}_{\nu}\sigma^{\nu\rho}F_{\nu\rho}\psi_{\nu}.
\end{eqnarray}

In the center-of-mass frame, we get the differential cross section \cite{hklyoon}:
\begin{eqnarray}
\left(\frac{d \sigma }{d \Omega }\right)_{D} &=& \frac{Q^2\alpha
\mu^2}{16 \pi} \sqrt{\frac{1- \frac{m_{\nu}^2}{E^2}}{1-
\frac{m_{q}^2}{E^2}}} \left[ \right. 1+\frac{m_{\nu}^2}{E^2}+\frac{m_{q}^2
m_{\nu}^2}{E^4} \nonumber \\
&& - \left( 1-\frac{m_q^2}{E^2}\right)\left(1-
\frac{m_{\nu}^2}{E^2}\right)
\cos^2{\theta} \left. \right]\label{dsigma}
\end{eqnarray}
and the total cross section
\begin{eqnarray}
 \sigma_{D}
 =  \frac{1}{6} Q^2\alpha \mu^2 \sqrt{\frac{1- \frac{m_{\nu}^2}{E^2}}{1- \frac{m_{q}^2}{E^2}}}
 \left(1+ \frac{m_q^2}{2E^2} \right) \left(1+ \frac{2m_{\nu}^2}{E^2}
 \right). \label{sigmat}
\end{eqnarray}
Here, D denotes Dirac type neutrinos.
A similar calculation  has been done by  Barut {\it et al.} \cite{barut}, where the total cross section is calculated as
\begin{eqnarray}
\sigma_{B}&=& \frac{ Q^2 \alpha \kappa^2 \sqrt{1- \frac{m_{\nu}^2}{E^2}}}{6 \sqrt{1 - \frac{m_{e}^2}{E^2}}} \left( 1 + \frac{1}{2} \frac{m_{e}^2}{E^2}
\right) \left( 1 + \frac{1}{2} \frac{m_{\nu}^2}{E^2} \right),
\end{eqnarray}
where $\kappa$ is the neutrino magnetic moment in their convention.  There is a small difference in the last term, which might be due to the additional $\left(1+\gamma_{5}\right)$ term in their interaction Lagrangian whereas we  consider  the case without chirality  for the Pauli coupling.  At high energy, $m_{\nu}$, $m_{q}$, $m_{e}\ll E$, which is the scale at the LHC or the UHECR of our interest,  they give the same result modulo the coupling constants:
\begin{eqnarray}
\left( \frac{d \sigma}{d \Omega} \right)_i =  \eta_i~ \sin^2
\theta, ~~
 \sigma_i
 =  \frac{8 \pi}{3} \eta_{i}, \label{sigmath}
\end{eqnarray}
where $i$'s stand for the choice of coupling constants, $\eta_D = Q^2 \alpha \mu^2/ 16 \pi $, and  $\eta_B = Q^2 \alpha \kappa^2/ 8 \pi$.
The cross section  becomes  constant  at high energy. If these are valid all the way to higher energy, then there is the problem of violation of the unitary bound.  However, the Pauli coupling is an effective interaction term that is valid only up to some scale,  and  we assume it to  be higher than the scale we are considering in this work.  Now, it is interesting to note that energy dependence is  quite different from that of pair production through the Z-boson channel in the standard model, where the cross section decreases with increasing colliding energy. The comparison and the possible implication will be   discussed in detail
in the final section.

\section{Pair production cross section of Majorana neutrinos }
The Majorana field is basically represented by a two-component spinor, $\chi$. For a free particle,
the Lagrangian of the two-component Majorana field is given by
\begin{eqnarray}
\mathcal{L}_{M}=\chi^{\dagger}i\bar{\sigma} \cdot \partial \chi -\frac{m}{2}\left[\left(\chi^{C}\right)^{\dagger}\chi+\chi^{\dagger}\chi^{C}\right].
\end{eqnarray}
$\chi$ can be written as
\begin{eqnarray}
\chi &=& \sum_{s}\int \frac{d^{3}\vec{p}}{\left(2\pi\right)^{3/2}\left(2E_{p}\right)^{1/2}}\left[ \right. f\left(\vec{p},s\right)a\left(\vec{p},s\right) e^{-ip \cdot x} \nonumber \\
&&+g\left(\vec{p},s\right)a^{\dagger}\left(\vec{p},s\right)e^{ip \cdot x} \left. \right],
\end{eqnarray}
where
$f$ and $g$ are two-component spinors that satisfy
the equation of motion for Majorana field,
\begin{eqnarray}
i \bar{\sigma} \cdot \partial \chi - i m \sigma^{2} \chi^{\ast} = 0. \label{Majoranaeq}
\end{eqnarray}
Then, it is possible to construct a four-component  Majorana field:
\begin{eqnarray}
\Psi_{M}
    &=& \sum_{s}\int \frac{d^{3}\vec{p}}{\left(2\pi\right)^{3/2}\left(2E_{p}\right)^{1/2}}\left[u\left(\vec{p},s\right)a\left(\vec{p},s\right)
e^{-ip \cdot x} \right. \nonumber \\
&& \left. +v\left(\vec{p},s\right)a^{\dagger}\left(\vec{p},s\right)e^{ip \cdot x} \right],
\end{eqnarray}
where
\begin{eqnarray}
u \equiv \left( \begin{array}{c} f\left(\vec{p},s\right) \\ g^{C}\left(\vec{p},s\right) \end{array} \right),  ~~ v \equiv \left( \begin{array}{c} g\left(\vec{p},s\right) \\ f^{C}\left(\vec{p},s\right) \end{array} \right).
\end{eqnarray}
Now, the interaction Lagrangian for the Majorana neutrino with the Pauli interaction can be written as
\begin{eqnarray}
\mathcal{L}_{int} &=& i \frac{\mu^{ij}}{2} \bar{\Psi}^{i}_{M} \sigma_{\mu\nu} \Psi^{j}_{M} F^{\mu \nu}, \label{LagM}
\end{eqnarray}
where $\sigma_{\mu\nu}=\frac{i}{2}\left[\gamma_{\mu},\gamma_{\nu}\right]$ and $g_{\mu\nu} = \left(+,-,-,-\right)$.  $\mu^{ij}$ is a transition magnetic moment that is antisymmetric for a Majorana neutrino, $\mu_{ij}=-\mu_{ji}$.

The differential cross section in the center of mass (CM) can be calculated in a straight forward way:   \begin{eqnarray}\label{e51-1}
&&\left( \frac{d\sigma}{d\Omega} \right)_{M} =\frac{\alpha Q^2\mu^2_{12}}{4\pi}\sqrt{\frac{E^2_1-m^2_{\nu_1}}{E^2-m^2_q}} \\
&& \times \left[ \frac{E_1 E_2}{E^2} + \frac{E_1 E_2 m^2_q}{2E^4} + \frac{m_{\nu_1}m_{\nu_2}}{E^2} + \frac{m^2_q m_{\nu_1}m_{\nu_2}}{2E^4} \right. \nonumber\\
&& \left. \quad - \frac{E_1 E_2 m^2_q \sqrt{1-\frac{m^2_{\nu_1}}{E^2_1}}\sqrt{1-\frac{m^2_{\nu_2}}{E^2_2}}}{2E^4} \right. \nonumber \\
&& \left. \quad -\frac{E_1 E_2}{E^2}\mathrm{ cos^2 \, \theta} \left(1-\frac{m^2_q}{E^2}\right)\sqrt{1-\frac{m^2_{\nu_1}}{E^2_1}}\sqrt{1-\frac{m^2_{\nu_2}}{E^2_2}} \right],\nonumber
\end{eqnarray}
where $\alpha$ is a fine-structure constant and  $M$ denotes the Majorana neutrino. After integrating over $d\Omega$,
the total cross section is given by
\begin{eqnarray}\label{e52}
&& \sigma_{M} = \frac{\alpha Q^2 \mu^2_{12}}{6}\frac{E_1\left(2+\frac{m^2_q}{E^2}\right)}{E\left(1-\frac{m^2_q}{E^2}\right)} \sqrt{1-\frac{m^2_{\nu_1}}{E^2_1}} \nonumber \\
&& \times ( \frac{3E_1 E_2}{E^2}+ \frac{3m_{\nu_1}m_{\nu_2}}{E^2} \nonumber \\
&& -\frac{E_1 E_2}{E^2} \sqrt{1-\frac{m^2_{\nu_1}}{E^2_1}}\sqrt{1-\frac{m^2_{\nu_2}}{E^2_2}} ).
\end{eqnarray}
In the high-energy region, where  the particle masses are very small compared
 to the energy scale, $m_i \ll E$,  the differential cross section and the total cross section, respectively, converge to simple expressions:
\begin{eqnarray}
\left(\frac{d\sigma}{d\Omega}\right)_M&=&\frac{\alpha Q^2\mu^2_{12}}{4\pi} \mathrm{ sin^2 \, \theta}\label{e53}
\end{eqnarray}
and
\begin{eqnarray}
\sigma_M = \frac{2\alpha Q^2 \mu^2_{12}}{3},\label{e54}
\end{eqnarray}
which are similar to the results for a Dirac neutrino with magnetic moment.

For comparison with
the neutrino pair production in the SM,
 in the high-energy limit $E \gg m_{\nu},M_{Z} $, the  differential cross section and the total cross section in the standard model are known to behave as
\begin{eqnarray}
\left( \frac{d\sigma}{d\Omega} \right)_{SM} \propto  (1+\cos^{2}
\theta), ~~
\sigma_{SM} \propto  \left(\frac{1}{E} \right)^2.
\end{eqnarray}
We can see that the total cross section behaves as
$\frac{1}{E^{2}}$ in the high-energy limit.  The angular distribution of the differential cross section is  maximum for  $\theta = 0$ and $\pi$  and minimum
for $\theta  \sim  \pi/2$.  These features are quite different from those  with the Pauli interaction, Eqs. (\ref{e53}) and (\ref{e54}).

\section{Discussion}
We  calculate the high-energy behavior of the cross section for Majorana neutrino pair production, assuming that the neutrinos have non-vanishing  transition magnetic moments and are interacting electromagnetically with the Pauli interaction.
 we found that the production cross section for Majorana-type neutrino production is similar to that of Dirac-type neutrino production.   The angular distribution in the center-of-mass frame  peaks at $\theta = \pi/2$ while the angular distribution in the standard model has a minimum at $\theta \sim \pi/2$.   The total  cross section turns out to be  independent of energy:
\begin{eqnarray}
\sigma_{M} & = & 1.66 \times 10^{-29} Q^{2} m^2 (\tilde{\mu})^2,
 \label{sigmatn} \end{eqnarray}
where  $\mu_{12} \equiv \tilde{\mu} \mu_B$. This is basically because the
neutrino interaction vertex carries a momentum factor of the
virtual photon, which cancels the energy dependence, which
is otherwise  inversely proportional to the square of the energy in the center-of-mass
frame as is the case for the standard model.
Hence, we can expect the neutrino production through the Pauli interaction to compete with that of the SM for high-energy collisions.  The energy scale, $E_{0.1}$, for which $\sigma_{M}$ becomes $0.1 ~ \sigma_{SM}$, can be estimated as
\begin{eqnarray}
E_{0.1} \sim 10^2 \left(\frac{10^{-10}}{\tilde{\mu}}\right) TeV.
\end{eqnarray}
For the Majorana-type neutrino, the upper bound is somewhat less stringent (although model dependent) than it is for the Dirac neutrino.  If we take  $\tilde{\mu} \lesssim 10^{-9} - 10^{-10}$\cite{numu},  we get
$E_{0.1} \sim 10 - 100$ $TeV$, which can be  reached in LHC and in UHECR experiments.  However, if $\tilde{\mu} < 10^{-11}$, the corresponding energy scale becomes  higher, beyond the GZK cutoff.
One of the characteristics of neutrino production by Pauli coupling is that the differential cross section has a maximum value $\theta = \pi/2$ in the center-of-mass frame, compared to the SM, which predicts a minimum at $\theta = \pi/2$.

Since the magnetic moment for a Majorana neutrino in this process
is not diagonal and can have only a transition magnetic moment, if  the neutrinos produced are  Majorana  type, then pairs
 should be produced with different flavors.  This difference gives us an additional way to find out  which type of neutrino is produced,   Majorana or Dirac.  However, it should be noted that most of the present experimental detector systems are such that the neutrinos produced in high-energy collisions escape detection.   Hence, for this purpose, we need detecting systems dedicated to high-energy neutrinos, for example, in high-energy cosmic ray experiments.

In summary, we discuss an observational possibility of a neutrino magnetic moment at high-energy  experiments and/or high-energy cosmic-ray experiments.  Although the present energy scale is found not to be sufficiently high enough for magnetic moments smaller than $\sim 10^{-11} \mu_B$, the neutrino magnetic moment with Pauli coupling  can open an interesting channel in future experiments, through which  the type  of neutrino can be distinguished.

\begin{acknowledgments}
The authors would like to thank Byung-Gu Cheon for useful discussions.   This work is supported by the
 World Class University (WCU) project of the Korean Ministry of Education, Science, and Technology (R33-2008-000-10087-0).
\end{acknowledgments}


\begin{thebibliography}{}
\bibitem{note} $E^{GZK}_{CM}$ is  the center of mass energy of UHECR ( $E^{GZK}\sim 10^{19}$ eV) and a proton  at rest ($E_p=m_p$),  $\sqrt{2 E^{GZK}m_p} \sim 100$ TeV.
\end{thebibliography}
\end{document}